\documentclass{emulateapj} 
\usepackage{apjfonts}
\journalinfo{The Astrophysical Journal, 2010, in press}

\shorttitle{ACCRETION AND CORONA IN TW HYA}
\shortauthors{BRICKHOUSE ET AL.}

\begin{document}

\title{A Deep Chandra X-ray Spectrum of the Accreting Young Star TW Hydrae}

\author{N. S. Brickhouse, S. R. Cranmer, A. K. Dupree, G. J. M. Luna,
  and S. Wolk}
\affil{Harvard-Smithsonian Center for Astrophysics, 60 Garden Street, Cambridge, MA 02138}

\newcommand{\xmm}{{\sl XMM-Newton}}
\newcommand{\Chandra}{{\it Chandra}}
\newcommand{\XMM}{{\sl XMM-Newton}}
\newcommand{\Sherpa}{{\it Sherpa}}
\newcommand{\sherpa}{{\it Sherpa}}

\begin{abstract}
We present X-ray spectral analysis of the accreting young star TW Hydrae
from a 489~ks observation using the  {\it Chandra} High Energy 
Transmission Grating. The spectrum provides a rich set of diagnostics 
for electron temperature $T_e$, electron density $N_e$, hydrogen column
density $N_H$, relative elemental abundances and velocities and
reveals its source in  3 distinct regions of the stellar atmosphere: 
the stellar corona, the accretion shock, and a very large extended 
volume of warm postshock plasma.  The presence of \ion{Mg}{12}, 
\ion{Si}{13}, and \ion{Si}{14} emission lines in the spectrum requires 
coronal structures at $\sim$10 MK. 
Lower temperature lines (e.g., from \ion{O}{8},
\ion{Ne}{9}, and \ion{Mg}{11}) formed at 2.5~MK appear more 
consistent with emission from an accretion shock.  He-like \ion{Ne}{9} 
line ratio diagnostics indicate that $T_e = 2.50 \pm 0.25$MK and
$N_e= 3.0\pm 0.2 \times 10^{12}$ cm$^{-3}$ in the shock.  These values
agree well with standard magnetic accretion models. However, the
{\it Chandra} observations significantly diverge from current model
predictions for the postshock plasma. This gas is expected to cool
radiatively, producing \ion{O}{7} as it flows into an increasingly 
dense stellar atmosphere.  Surprisingly, \ion{O}{7} indicates
$N_e=5.7^{+4.4} _{-1.2}\times 10^{11}$ cm$^{-3}$, five times lower than 
$N_e$ in the accretion shock itself, and $\sim$ seven times lower than
the model prediction.  We estimate that the postshock region
producing \ion{O}{7} has roughly 300 times larger volume, and 30 times more 
emitting mass than the shock itself.  Apparently, the shocked plasma
heats the surrounding stellar atmosphere to soft X-ray emitting temperatures
and supplies this material to nearby large magnetic 
structures -- which may be closed magnetic loops or open
magnetic  field leading to mass outflow.  Our model explains
the soft X-ray excess found in many accreting systems as well as the
failure to observe high $N_e$ signatures in some stars. 
Such accretion-fed coronae may be ubiquitous in the
atmospheres of accreting young stars.
\end{abstract}


\section{Introduction}

Low mass stars in star-forming regions produce strong X-ray emission
from coronal magnetic activity, as evidenced by high temperature
($\sim$10~MK) emission from flares and active regions (Feigelson \&
Montmerle 1999; Gagn\'{e} et al.\  2004; Preibisch et al.\  2005). The
role of accretion in the production of X-rays from young stars is less
well understood.  Magnetic structures on stars can extend all the way
to the inner accretion disk, as suggested from models of decaying
stellar flares (Favata et al.\  2005). Extended X-ray emission from jets
also occurs, as observed in DG~Tau, a system oriented so that the
circumstellar disk plane aligns with our line of sight (G\"{u}del et
al. 2008).  Soft X-ray emission may arise from shocks at the stellar
base of this extended X-ray jet (G\"{u}del et al.\  2008; Schneider \&
Schmitt 2008).

X-rays can also be produced in accreting systems when the accreting
material accelerates to supersonic velocities and shocks near the
stellar surface, heating the gas to a few MK. In the standard model of
accreting Classical T Tauri Stars (CTTS), the magnetic field of the
star connects to the accretion disk near the corotation radius,
channeling the accretion stream from the disk to a small area on the
star (Calvet \& Gullbring 1998). X-rays from the shock may be difficult to
detect if the shock is formed too deep in the photosphere (Hartmann 1998) or
if they are absorbed by the stream of preshocked neutral or near-neutral
gas. Furthermore, the X-ray shock signature may be difficult to distinguish from
soft coronal emission without the high resolution spectroscopy
necessary to determine temperature and density and identify the
emitting regions. Shock models predict
high electron density $N_e$ ($\sim 10^{13}$ cm$^{-3}$) at relatively
low electron temperature $T_e$ (a few MK), features that can be tested
with X-ray line ratio diagnostics.

TW~Hydrae (TW~Hya) was the first CTTS found to show these
characteristics of high density and low temperature (Kastner et
al. 2002; hereafter, K02), based on a 48~ks \Chandra\ High Energy
Transmission Grating (HETG) spectrum. TW Hya is one
of the oldest known stars ($\sim$~10 Myr) still in the CTTS (accreting) phase. It 
is uniquely poised in an interesting state of evolution when it will
soon stop accreting, lose its disk, and perhaps form planets
(e.g., Calvet et al.\  2002). At a distance of only 57 pc, it is also one of the
brightest T Tauri stars in X-rays. Interstellar absorption is
minimal, and the observed absorption can be considered intrinsic to
the system (K02). TW Hya's circumstellar disk is nearly face-on with
an inclination angle of  7$^{\rm{o}}$  (Qi et
al. 2004). Thus the star appears pole-on.

Key to K02's argument for accretion is the
electron density ($N_e \sim 6 \times 10^{12}$ cm$^{-3}$) determined
from He-like \ion{Ne}{9} emission line ratios, and confirmed by 
subsequent observations with the \xmm\ Reflection Grating Spectrometer
(Stelzer \& Schmitt 2004) and the \Chandra\ Low Energy Transmission
Grating (LETG; Raassen 2009). While $N_e$ for some active cool star
coronae approaches such high values at high $T_e$ ($\sim 10$ MK), 
e.g., 44~Boo, UX~Ari, and II~Peg (Sanz-Forcada et al.\  2003; Testa et al.\  2004), the high $N_e$ of
TW~Hya is produced at the significantly lower $T_e$ ($\sim$3~MK)
expected for the accretion shock.  BP Tau (Schmitt et al.\  2005),
V4046~Sag (Gunther et al.\  2006), RU~Lup (Robrade \& Schmitt 2007), and
MP Mus (Argiroffi et al.\  2007) also show high values of the electron
density. On the other hand, T~Tau and AB~Aur have a low $N_e$
(G\"{u}del et al.\  2007a; Robrade \& Schmitt 2007); and, the binary
Hen~3-600 appears to lie at an intermediate value of $N_e$
(Huenemoerder et al.\  2007). Despite large differences in $N_e$, all of
these stars show an excess ratio of soft X-ray emission to hard X-ray
emission, manifested by their large \ion{O}{7}/\ion{O}{8} ratios
as compared with active main sequence stars (G\"{u}del \& Telleschi
2007; Robrade \& Schmitt 2007). Additional soft X-ray diagnostics available with a deep
\Chandra\ exposure might be expected to shed light on the accretion
process and test the shock models.

Since accretion is expected to result in higher densities than found in an
active stellar corona, it is critical that the density determination be
secure. Interpretation of the \ion{Ne}{9} line ratio
in terms of high density is not unique, since photoexcitation by
ultraviolet radiation mimics the effects of high density. Determining
the density from \ion{Mg}{11} is not subject to the same degeneracy, since the
relevant ultraviolet radiation, observed with the Far Ultraviolet
Explorer (FUSE; Dupree et al.\  2005), is too weak to photoexcite its
metastable level. The three He-like systems (\ion{O}{7}, \ion{Ne}{9},
and \ion{Mg}{11}) together can provide independent determinations not
only of the electron density, but also of the electron temperature and
hydrogen column density. Deep spectroscopy can, first, put the
accretion interpretation on firm ground and second, allow us to
determine the structure of the accretion
region itself.

We were awarded a \Chandra\ Large Observing Project to establish
definitively whether the X-ray emission in TW~Hya is associated with
accretion, using high signal-to-noise ratio spectroscopy with the
HETG, and, if so, to determine the emission measure distribution and
elemental abundances, search for X-ray signatures of infall, outflow,
and turbulence, and constrain the properties of the accretion hot
spot. During the \Chandra\ observations we also conducted a
ground-based observing campaign, which will be reported elsewhere
(Dupree et al.\  2010, in preparation). Here, we describe the X-ray
observations in Section~2 and report analysis results in Section~3,
where the densities, temperatures, and column densities can be
extracted from this rich spectrum. Section~3 also presents elemental
abundances derived from fits to the spectrum, as well as the
measurement of turbulent velocity. Section~4 describes spectral
predictions from a model of the TW~Hya shock.  In Section~5 we discuss
evidence for a new type of coronal structure. Section~6 gives a
summary of results and conclusions.

\section{Observations and Spectral Analysis}

\Chandra\ observed TW~Hya with the HETG in combination with the ACIS-S
detector array for a total of 489.5~ks intermittently spanning the
period between 2007 Feb 15 and 2007 Mar 3. The observation consists of 4 segments, with
exposure times as follows: 153.3~ks (Obsid 7435), 157.0~ks (Obsid 7437), 158.4~ks (Obsid 7436),
and 20.7~ks (Obsid 7438). The summed spectrum has about ten times the
exposure time of the spectrum analyzed by K02. Figure~1 shows the total first-order light
curve, with flux variations of about a factor two during the
observation. Assuming a distance of 57 pc, the observed X-ray luminosity $L_X$, measured using the
Medium Energy Grating
between 2.0 and 27.5 \AA, is $1.3 \times 10^{30}$ erg s$^{-1}$.  One
clear flare, with total X-ray luminosity of 2.1 $\times 10^{30}$ erg s$^{-1}$,
persists for $\sim$15~ks, producing a total energy of 4 $\times
10^{34}$ ergs. The flare is retained in the following analysis since
its contribution to line fluxes is negligible (see Section
3.4).

\begin{figure}
\epsscale{1.15}
\plotone{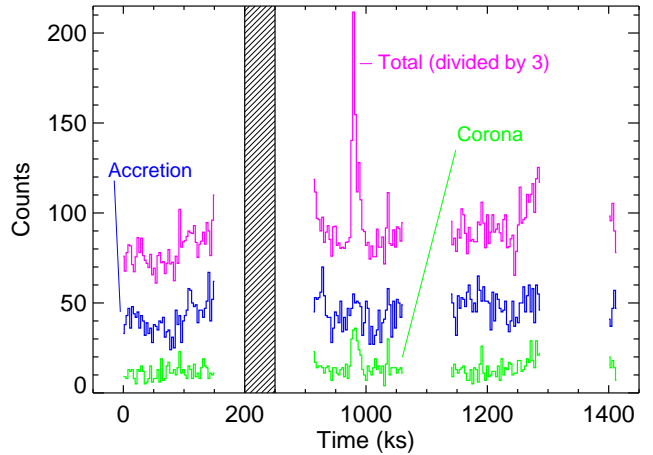}
\caption{X-ray light curves (first order counts vs time since start) for the four
  observing segments of the \Chandra\ observation. The hashed area
  represents a gap of 650~ks. The upper curve
  shows the light curve for the total first order counts. This light
  curve includes both line and continuum emission. The middle
  curve uses the combined line counts from \ion{N}{7}, \ion{O}{7},
  \ion{O}{8}, \ion{Ne}{9}, \ion{Fe}{17}, and \ion{Mg}{11} (identified
  as the
  ``accretion'' lines using Model D, discussed in Section 3.4). The lower curve uses the combined line counts
  from \ion{Mg}{12}, \ion{Si}{13}, and \ion{Si}{14} (identified as the
  lines from the ``corona'' from Model D).
  Note that the line-based light
  curves also include any continuum counts within the line profile. The data are binned over 3~ks.
}
\end{figure}

We have extracted spectra from the event files, and
produced calibration files using the \Chandra\ Interactive Analysis of
Observations (CIAO ver. 3.4) software package (Fruscione et al.\  2006).
With response matrices combined from the four observations, \sherpa\
(Freeman et al.\  2001) was used to extract individual emission line
fluxes from the co-added observations, maintaining separate first
order spectra
for the four grating arms (plus and minus; MEG and HEG) during the
fitting procedure. A model for the continuum plus background was found from a global fit to
line-free regions and used in addition to the line models. Background
rejection is high with ACIS order sorting. Thus the background
contributes relatively little to this model except at the long
wavelength end of the spectrum, where the effective area is low. In
the O VII region, for example, the background contains about half as
many counts as the continuum. For the relatively weak but important
\ion{O}{7} line, background, continuum, and line have about 1,
2, and 16 counts, respectively. 

Line models
are Gaussian functions with negligible or small widths relative to the
width of the instrumental line response functions ($\sim$0.012~\AA\ and
0.023~\AA, for HEG and MEG, respectively). Table~1
gives the identified lines and their observed fluxes. For most lines the fit
widths are less than half the instrumental width and are not
statistically significant.  We discuss the
widths of the strongest lines in Section~3.5.

\begin{deluxetable}{lrccl}
\tabletypesize{\footnotesize}
\tablewidth{0pc}
\tablecaption{Selected Emission Lines}
\tablehead{
\colhead{Line} & 
\colhead{$\lambda_{ref}$\tablenotemark{a}} &
\colhead{Flux\tablenotemark{b}} & 
\colhead{Eff Area\tablenotemark{c}} &
\colhead{Origin\tablenotemark{d}} \\
\colhead{} & 
\colhead{(\AA)} &
\colhead{(10$^{-6}$ ph cm$^{-2}$ s$^{-1}$)} &
\colhead{(cm$^2$)} &
\colhead{} 
} 
\startdata
\ion{Si}{14} &  6.182   & 1.87 $\pm$ 0.19 &  163.0 & corona \\
\ion{Si}{13} &  6.648   & 2.72 $\pm$ 0.17 &  146.6 & corona \\
\ion{Si}{13} &  6.688   & 0.74 $\pm$ 0.14 &  138.3 & corona \\
\ion{Si}{13} &  6.740   & 1.66 $\pm$ 0.15 &  151.6 & corona \\
\ion{Mg}{12} &  8.421   & 2.24 $\pm$ 0.20 &  162.4 & corona \\
\ion{Mg}{11} &  9.169   & 2.25 $\pm$ 0.22 &  135.4 & shock \\
\ion{Mg}{11} &  9.231   & 0.90 $\pm$ 0.17 &  133.2 & shock \\
\ion{Mg}{11} &  9.314   & 1.27 $\pm$ 0.19 &  130.1 & shock \\
\ion{Ne}{10} &  9.481   & 1.64 $\pm$ 0.20 &  114.5 & corona \\
\ion{Ne}{10} &  9.708   & 2.75 $\pm$ 0.27 &  86.2 & corona \\
\ion{Ne}{10} & 10.239   & 9.23 $\pm$ 0.53 &  81.1 & corona \\
\ion{Ne}{ 9}\tablenotemark{e} & 10.765   & 3.54 $\pm$ 0.43 & 67.9 & shock \\ 
\ion{Ne}{ 9} & 11.001   & 8.48 $\pm$ 0.63 &  67.4 & shock \\
\ion{Ne}{ 9} & 11.544   & 23.7 $\pm$ 0.99 &  55.3 & shock \\
\ion{Fe}{22} & 11.783   & 3.28 $\pm$ 0.50 &  52.3 & corona \\
\ion{Fe}{22} & 11.932   & 3.01 $\pm$ 0.50 &  49.6 & corona \\
\ion{Ne}{10} & 12.134   & 76.1 $\pm$ 1.9  &  48.5 & corona \\
\ion{Fe}{17} & 12.266   & 4.3  $\pm$ 0.6  &  46.6 & mixed\tablenotemark{f} \\
\ion{Ne}{ 9} & 13.447   & 177. $\pm$ 4.2  &  22.3 & shock \\
\ion{Ne}{ 9} & 13.553   & 114. $\pm$ 3.3  &  22.7 & shock \\
\ion{Ne}{ 9} & 13.699   & 58.7 $\pm$ 2.3  &  25.0 & shock  \\
\ion{Fe}{18} & 14.208   & 4.59 $\pm$ 0.76 & 29.0 & corona \\
\ion{Fe}{18} & 14.256   & 1.63 $\pm$ 0.53 & 28.5 & corona \\
\ion{O }{ 8} & 14.820   & 5.04 $\pm$ 0.80 & 23.7 & shock \\
\ion{Fe}{17} & 15.014   & 36.5 $\pm$ 1.8  & 21.9 & shock \\
\ion{O }{ 8} & 15.176   & 10.4 $\pm$ 1.2  & 20.8 & shock \\
\ion{Fe}{17} & 15.261   & 17.4 $\pm$ 1.4  & 20.0 & shock \\
\ion{O }{ 8}\tablenotemark{g} & 16.006   & 30.6 $\pm$ 2.3 & 15.9 & shock\\
\ion{Fe}{18} & 16.071   & 5.17 $\pm$ 1.0 & 16.3 & corona \\
\ion{Fe}{17} & 16.780   & 25.5 $\pm$ 3.0 & 12.1 & shock \\
\ion{Fe}{17} & 17.051   & 27.9 $\pm$ 2.3 & 11.1 & shock  \\
\ion{Fe}{17} & 17.096   & 26.3 $\pm$ 2.3 & 11.2 & shock \\
\ion{O }{ 7} & 17.207   & 4.68 $\pm$ 1.4 &  10.7 & shock \\
\ion{O }{ 7} & 17.396   & 4.68 $\pm$ 1.5 &  9.8 & shock \\
\ion{O }{ 7} & 17.768   & 3.66 $\pm$ 1.6 &  7.7 & shock \\
\ion{O }{ 7} & 18.627   & 17.9 $\pm$ 1.4 &  7.3 & shock \\
\ion{O }{ 8} & 18.969   & 213. $\pm$ 8.4 &  6.6 & shock \\
\ion{N }{ 7} & 20.910   & 3.8  $\pm$ 2.5 &  3.3 & shock \\
\ion{O }{ 7} & 21.601   & 117. $\pm$ 10.0 & 2.8 & postshock\tablenotemark{h}   \\
\ion{O }{ 7} & 21.804   & 72.4 $\pm$ 9.1  & 2.7 & postshock\tablenotemark{h} \\
\ion{O }{ 7} & 22.098   & 15.2 $\pm$ 4.4  & 2.1 & postshock\tablenotemark{h} \\
\ion{N }{ 7} & 24.781   & 68.5 $\pm$ 7.6  & 2.6 & shock \\
\enddata
\tablenotetext{a}{Reference wavelength from ATOMDB. Wavelengths for multiplets are given as intensity-weighted averages.}
\tablenotetext{b}{Observed fluxes at Earth, with no correction for
  absorption. Flux errors are 1$\sigma$.}
\tablenotetext{c}{Effective areas are estimate by combining the four grating arms and
  averaging the effective areas  over the line width.}
\tablenotetext{d}{The origin of the emission lines is based on Model~D
  parameters (Section 3.4).}
\tablenotetext{e}{Blend with \ion{Fe}{17} $\lambda$10.77.}
\tablenotetext{f}{It is interesting to note that this \ion{Fe}{17}
line is mixed, while the longer wavelength lines are primarily from
the shock. The reason for this difference is that the 12.266 \AA\ line
flux is relatively stronger at higher temperature.}
\tablenotetext{g}{Blend with \ion{Fe}{18} $\lambda$16.004; however,
  the predicted \ion{Fe}{18} contribution is negligible.}
\tablenotetext{h}{This line is primarily formed in the postshock
  cooling plasma (Section~5). In Model~D the EM distribution has only
  one component representing both the shock and postshock regions. }
\end{deluxetable}

All line fluxes are within 3$\sigma$ of the values reported by K02
except for \ion{Ne}{9} $\lambda$13.437, which has slightly more than
twice as much flux in our observation. Our line fluxes are also within
3$\sigma$ of the values reported by Stelzer \& Schmitt (2004) for
lines observed with both instruments, except for \ion{Ne}{9}
$\lambda$13.553, for which our flux is about half. For lines in common
with those reported by Raassen (2009) all fluxes agree within
3$\sigma$. With the higher signal-to-noise ratio of our data, new
diagnostic information is available.  

A major goal of this study is the determination of electron density
from the He-like ions. We first discuss the diagnostics, and then
model the global spectrum. Figure~2 shows the high quality in the He-like \ion{O}{7} and
\ion{Ne}{9} line regions. Interestingly, the \ion{Ne}{9} region shows
no apparent \ion{Fe}{19} lines, which can contaminate the \ion{Ne}{9} diagnostics
(e.g., Ness et al.\  2003). Upper limits on the strongest \ion{Fe}{19}
line at 13.52 \AA\ imply that \ion{Fe}{19} contributes 1\% or less to the \ion{Ne}{9}
line fluxes. 

\begin{figure}
\includegraphics[width=0.45\textwidth]{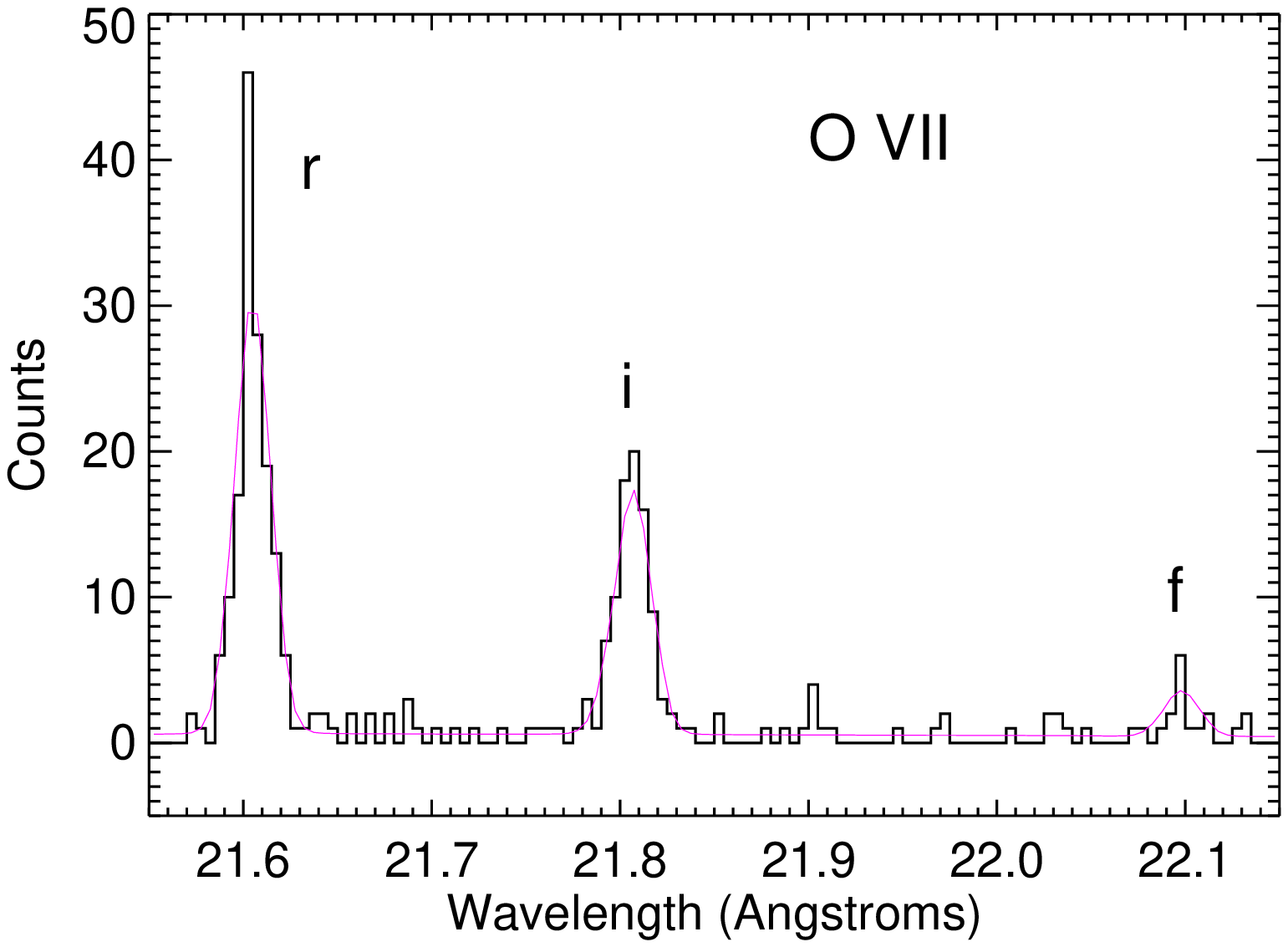}
\includegraphics[width=0.47\textwidth]{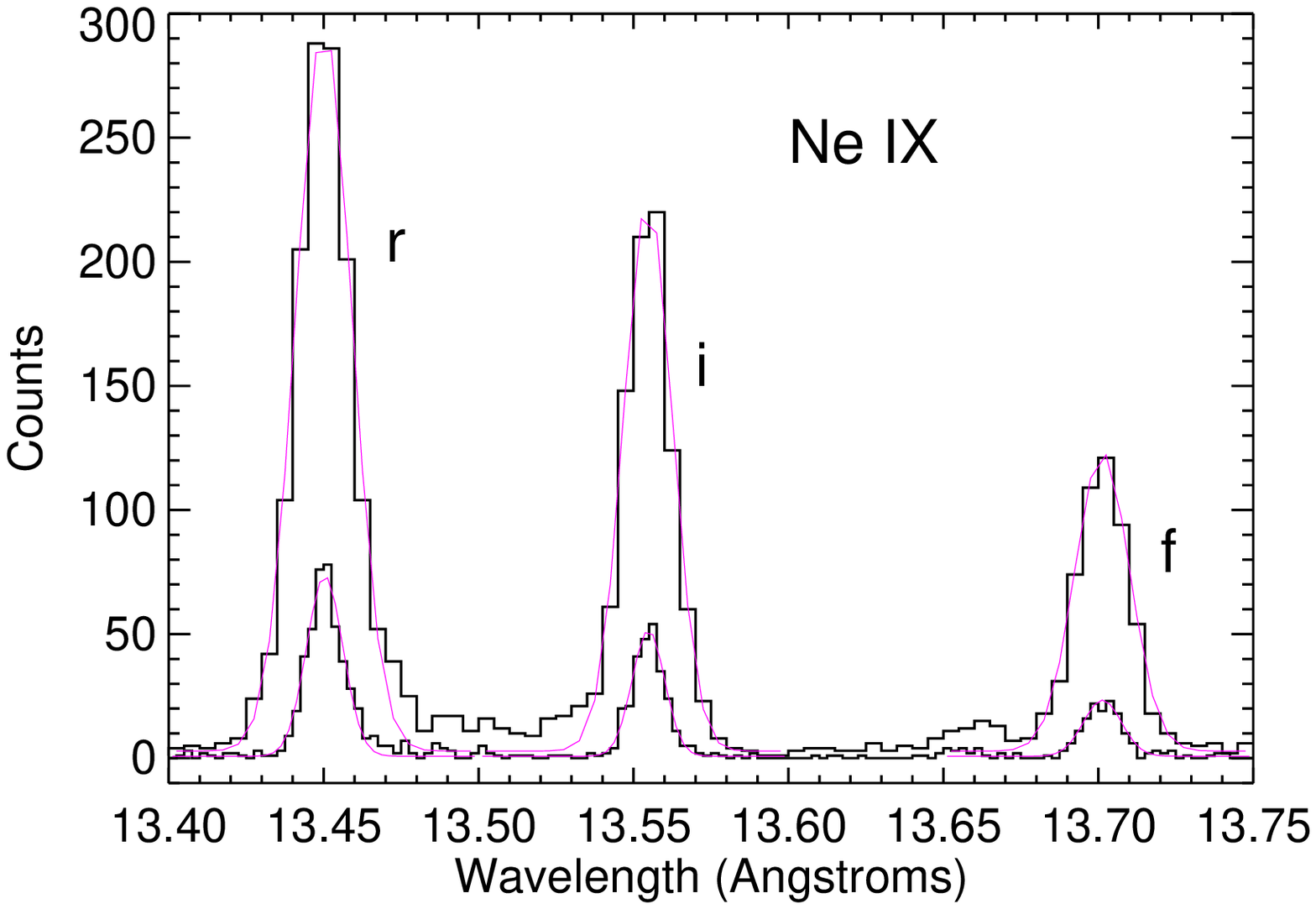}
\caption{{\it Top:} Observed MEG spectrum in the \ion{O}{7} spectral
region, shown as a histogram, with best-fit model (3 Gaussian lines)
overlaid.  Wavelength positions are fixed relative to each other and
the instrumental line response function is included in the model. Resonance (r),
intercombination (i) and forbidden (f) lines are marked. {\it Bottom:}
Observed MEG ({\it upper}) and HEG ({\it lower}) spectra in the
\ion{Ne}{9} region, shown as histograms, with best-fit model
overlaid. For this fit, the positions, fluxes and line widths are free
parameters, yielding the constraints on velocities described in the
text.}
\end{figure}

The \ion{Mg}{11} triplet line region is more
complex (Fig.~3), with lower signal-to-noise ratio and apparent
line blending between 9.2 and 9.4 \AA.  For this region, the fit is
constrained such that all known lines are fixed in wavelength relative
to the \ion{Mg}{11} resonance line $\lambda$9.169, which is fit
separately for each grating arm.  Lines are Gaussian functions with
widths set to zero (i.e., only the instrumental line response function
is used). Line fluxes of the \ion{Ne}{10} Lyman series
lines $n=5$ to $10$ are scaled according to their oscillator
strengths, since collision strengths are not available in the
literature. We note that systematic errors from line blends are not 
included in the errors given in Table~1.

\section{Results}

One of the advantages of X-ray spectroscopy with high signal-to-noise characteristics
is the availability of numerous line ratio diagnostics to assess
conditions in the radiating plasma. In particular,
He-like systems provide valuable diagnostic opportunity, as
discussed extensively in the literature (Porquet et al.\  2001; Smith et
al.\  2001; and references therein). The G-ratio, defined as the flux
ratio of the forbidden ($1s2s~^3S_{1}  \rightarrow 1s^2~^1S_{0}$) plus
intercombination ($1s2p~^3P_{1,2} \rightarrow 1s^2~^1S_{0}$) lines to the
resonance ($1s2p~^1P{_1} \rightarrow 1s^2~^1S_{0}$) line, is sensitive to the
electron temperature ($T_e$), while the R-ratio (flux ratio of the
forbidden to intercombination lines) is sensitive to $N_e$. Ratios of
emission lines in the resonance line series (transitions from 2p, 3p, etc.  to
ground, also known as He$_\alpha$\footnote{He$_\alpha$ is the
resonance line sometimes labeled $r$ or $w$.}, He$_\beta$, and so on)
are sensitive both to $T_e$, through their relative Boltzmann factors,
and to the absorption by an intervening  hydrogen column density ($N_H$) through their energy
separation.

\begin{figure}
\epsscale{1.09}
\plotone{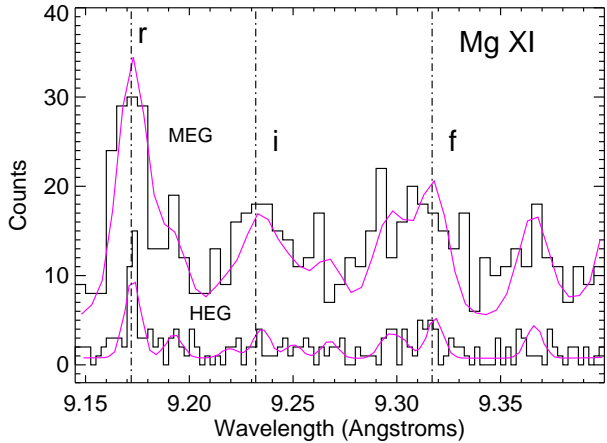}
\caption{Observed MEG ({\it upper}) and HEG ({\it lower}) spectra in
the \ion{Mg}{11} spectral region, shown as histograms, with best-fit
model spectrum, described in text, overlaid. Positions of resonance
(r), intercombination (i), and forbidden (f) lines are marked.}
\end{figure}

This analysis emphasizes the use of line ratio diagnostics, including
those discussed above.  We use the
Astrophysical Plasma Emission Code (APEC) with the atomic data in
ATOMDB v1.3\footnote{http://cxc.harvard.edu/atomdb} (Smith et al.\  2001)
to calculate line emissivities, except
for He-like \ion{Ne}{9}, for which we use the more accurate
calculations of Chen et al.\  (2006). The model for absorption by gas
with cosmic abundances is taken from Morrison \& McCammon (1983).

In this section we present the physical conditions of the emitting
plasma as determined from the line ratios of specific He-like and
other ions, in particular $N_e$, $T_e$, and $N_H$. We use information
determined from line ratios to constrain an empirical model of the
emission measure distribution (EMD), and then allow additional
parameters of the model to be constrained by global fits to the
spectrum. We give the elemental abundances determined from these
fits.  Finally we present velocity constraints from line centroids
and widths.

\subsection{Electron temperature}

We use G-ratios from multiple ions to determine the electron
temperature. Figure~4 shows the observed G-ratios of He-like O, Ne and Mg,
overplotted on their respective theoretical functions of $T_e$. For
comparison, the emissivities of their resonance lines peak at $T_e =$
2, 4, and 6~MK, respectively, as shown in the figure. Table~2 gives
the derived $T_e$ from these ions.  The \ion{Ne}{9} G-ratio indicates
that $T_e = 2.5 \pm 0.25 $ MK. The \ion{Ne}{9} $T_e$ determination is
highly reliable, given that recent theoretical calculations, which
include the resonance contributions to the collisional excitation
cross sections, agree with experimental measurements to within 7\%
(Chen et al.\  2006; Smith et al.\  2009a, 2009b).  On the other hand, the
theoretical calculations for \ion{O}{7} and \ion{Mg}{11} have higher
systematic uncertainties, and the observational errors are larger,
such that the $T_e$ values derived from \ion{O}{7} and \ion{Mg}{11}
carry less weight (but are consistent with \ion{Ne}{9}). Both
\ion{Ne}{9} and \ion{Mg}{11} are found far below their temperature of
maximum emissivity $T_{max}$, supporting an accretion shock scenario
for their formation.

\begin{figure}
\epsscale{1.14}
\plotone{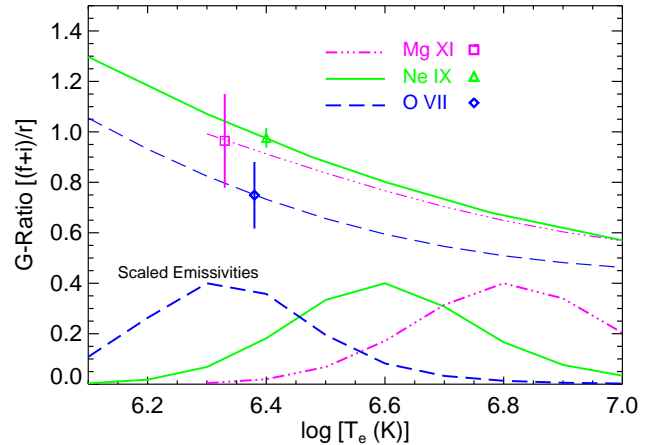}
\caption{Theoretical G-ratios (i.e., ratios of forbidden plus
intercombination to resonance line fluxes) as functions of $T_e$, for
\ion{O}{7} ({\it dashed}), \ion{Ne}{9} ({\it solid}), and \ion{Mg}{11}
({\it dash-dotted}). Overplotted are the observed ratios with 1$\sigma$
errors. The model for \ion{Ne}{9} is from Chen et al.\  (2006). The
models for \ion{O}{7} and \ion{Mg}{11} are from APEC using ATOMDB
v1.3 (Smith et al.\  2001). Scaled emissivities (ph cm$^{3}$ s$^{-1}$) for the resonance
lines of the three ions are
also shown. For \ion{Ne}{9} and \ion{Mg}{11} the derived temperatures
are well below the temperatures of their peak emissivities, consistent
with an intepretation of their formation in an accretion shock.}
\end{figure}

\begin{deluxetable}{lcc}
\tabletypesize{\footnotesize}
\tablewidth{0pc}
\tablecaption{He-like G-Ratio $T_e$ Diagnostics}
\tablehead{
\colhead{Ion} &
\colhead{Observed Ratio\tablenotemark{a}} &
\colhead{$T_e$} \\
\colhead{} &
\colhead{} &
\colhead{(K)}
}
\startdata
\ion{O}{7}  &    0.75$\pm$0.13 & 2.4$^{+1.2}_{-0.6} \times
10^6$  \\
\ion{Ne}{9} &  0.98$\pm$0.04 & 2.5$^{+0.25}_{-0.25} \times 10^6$ \\
\ion{Mg}{11} & 0.97$\pm$0.19 & 2.1$^{+2.1}_{-0.5} \times 10^6$
\enddata
\tablenotetext{a}{Errors are 1$\sigma$.}
\end{deluxetable}

\subsection{Electron density}

We use R-ratios from multiple ions to determine the electron
density. Figure~5 shows the observed R-ratios for \ion{O}{7},
\ion{Ne}{9}, and \ion{Mg}{11}, overplotted on the theoretical
functions of electron density $N_e$ from APEC. Table~3 presents the
$N_e$ values derived from these R-ratios.  Sensitivity to $T_e$ is
negligible near the observed R-ratio values. Smith et al.\  (2009) show
that the APEC R-ratio for \ion{Ne}{9} in the high density range is
indistinguishable from the R-ratio calculated by Chen et al.\  (2006). 
The theoretical R-ratios from APEC for all three ions are in excellent
agreement with Porquet et al.\  (2001) over the density range reported
here.

\begin{deluxetable}{lcc}
\tabletypesize{\footnotesize}
\tablewidth{0pc}
\tablecaption{He-like R-Ratio $N_e$ Diagnostics}
\tablehead{
\colhead{Ion} &
\colhead{Observed Ratio\tablenotemark{a}} &
\colhead{$N_e$} \\
\colhead{} &
\colhead{} &
\colhead{(cm$^{-3}$)}
}
\startdata
\ion{O}{7}  &    0.21$\pm$0.07 & 5.7$^{+4.4}_{-1.2} \times 10^{11}$  \\
\ion{Ne}{9} &  0.51$\pm$0.03 & 3.0$^{+0.2}_{-0.2} \times  10^{12}$ \\
\ion{Mg}{11} & 1.41$\pm$0.34 & 5.8$^{+3.8}_{-2.4} \times 10^{12}$
\enddata
\tablenotetext{a}{Errors are 1$\sigma$.}
\end{deluxetable}

From \ion{Ne}{9} we derive $N_e = 3.0 \pm 0.2\times 10^{12}$
cm$^{-3}$, while \ion{O}{7} gives $N_e = 5.7 ^{+4.4} _{-1.2} \times
10^{11}$ cm$^{-3}$.  The 3$\sigma$ range from \ion{Ne}{9}, between 2.5
and 3.9 $\times 10^{12}$ cm$^{-3}$, provides a tight
constraint.\footnote{The measured value of the \ion{Ne}{9} R-ratio is
0.515 $\pm$ 0.033, quite close to the value of 0.446 $\pm$ 0.124
reported by K02. The difference of nearly a factor of 2 in the derived
$N_e$ appears to be due to the difference in atomic data used, with
their models possibly based on the Raymond \& Smith code (Raymond
1988).} The best-fit value for \ion{Mg}{11}, using all the lines in
the region as described in Section 2, is $N_e = 5.8 ^{+3.8} _{-2.4}
\times 10^{12}$ cm$^{-3}$. For comparison, we have also fit only the 3
lines of Mg XI, again with line widths and relative positions fixed,
obtaining $N_e =$ 7 $\pm$ 2 $\times 10^{12}$ cm$^{-3}$, in good
agreement with the best-fit value. This comparison suggests that line
blending does not unduly affect the result.

\begin{figure}
\epsscale{1.10}
\plotone{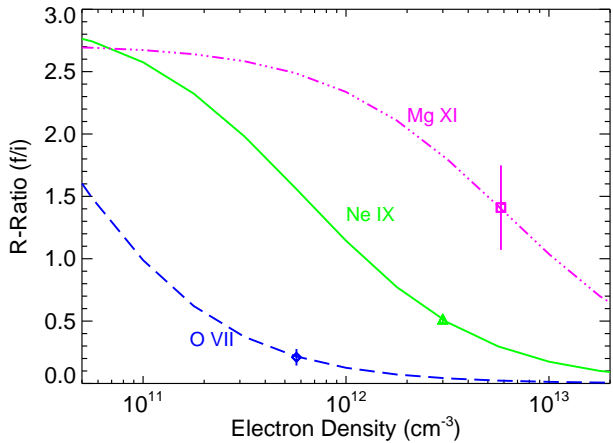}
\caption{Theoretical R-ratios (i.e., ratios of forbidden to
intercombination line fluxes) as functions of $N_e$ for He-like O, Ne
and Mg. Overplotted are the observed ratios with 1$\sigma$
errors. Models are from APEC using ATOMDB v1.3. While Figure~4 shows
that these ions are formed at similar temperature, they are found at
significantly different densities.}
\end{figure}

At low $N_e$ the population of the metastable $2s~^3S_1$ level builds
up since the decay rate is extremely low. At high $N_e$ electron
impact excites the metastable level population up to the $2p^3P$
levels. In principle, low R-ratios may also be produced at low $N_e$
by photoexcitation. Stelzer \& Schmitt (2004) and K02 dismiss
photoexcitation based on the stellar temperature ($\sim$ 4000~K), but
Drake (2005) suggests that a hot spot above the stellar surface, where
the accretion stream shocks, could produce sufficient photoexciting
radiation. We have measured the flux at the excitation wavelengths
1647 \AA, 1270 \AA\ and 1034 \AA\ for \ion{O}{7}, \ion{Ne}{9}, and
\ion{Mg}{11}, respectively, from the {\it Hubble Space Telescope} STIS
(Herczeg et al.\  2002) and the FUSE spectra
(Dupree et al.\  2005). The intensity at 1034 \AA\ is an order of
magnitude lower than at the longer wavelengths, constraining the black
body temperature of the hot spot to be less than 10,000~K, in good
agreement with {\it International Ultraviolet Explorer} analysis
showing a blackbody at 7900~K covering about 5\% of the stellar
surface (Costa et al.\  2000). Thus the ultraviolet measurements and the
\ion{Mg}{11} R-ratio rule out photoexcitation and support the
interpretation of high $N_e$. Together with the temperature derived
from the \ion{Ne}{9} G-ratio, the density supports an accretion
origin for \ion{Ne}{9}.

The ratio of \ion{Fe}{17} $\lambda$17.096 to $\lambda$17.051 becomes
sensitive to $N_e$ above $\sim 10^{13}$ cm$^{-3}$ (Mauche et
al. 2001).  The measured ratio of 1.06 $\pm$ 0.13 is at the low $N_e$
limit, ruling out the high density from this line ratio suggested by
Ness \& Schmitt (2005).

\subsection{Hydrogen column density}

The HETG spectrum contains several series of resonance lines that indicate
absorption by neutral gas along the line of sight.  The most
useful of these diagnostics are the \ion{O}{7} and \ion{Ne}{9}
He$_\alpha$/He$_\beta$ ratios because their formation temperatures are
independently determined from the G-ratio. Table~4 presents the observed
line ratios, and their use as diagnostics for $N_H$. Figure~6 shows the
theoretical He$_\alpha$/He$_\beta$ ratios as functions of $T_e$ for
\ion{O}{7} and \ion{Ne}{9}, for both unabsorbed and absorbed
cases. The observed ratios are overplotted using $T_e$ determined from
their G-ratios (which, as noted above, is more reliable for
\ion{Ne}{9} than for \ion{O}{7}). For the best-fit $T_e$, \ion{Ne}{9}
gives $N_H = 1.8 \pm 0.2 \times 10^{21}$ cm$^{-2}$ and \ion{O}{7}
gives $N_H = 4.1 ^{+1.9} _{-1.6} \times 10^{20}$ cm$^{-2}$. These two
values show a meaningful difference. For \ion{Ne}{9} at 2.5~MK, the
observed ratio is 11$\sigma$ below the unabsorbed model, making the
presence of an absorbing column quite definitive. The column density
from \ion{O}{7} is consistent with the lower limit of $N_H = 2 \times
10^{20}$ cm$^{-2}$ determined from the absence of \ion{Ne}{8} and
\ion{Si}{12} lines longward of 40 \AA\ in the \Chandra\ LETG spectrum (Raassen 2009).

\begin{deluxetable}{lcrrrr}
\tabletypesize{\footnotesize}
\tablewidth{0pc}
\tablecaption{Line Ratio $N_H$ Diagnostics}
\tablehead{
\colhead{Identification} &
\colhead{Observed} &
\multicolumn{4}{c}{Predicted Ratio for log $N_H$ (cm$^{-2}$):} \\
\colhead{} &
\colhead{Ratio\tablenotemark{a}} &
\colhead{0.00} &
\colhead{20.61 } &
\colhead{20.89} &
\colhead{21.26}
}
\startdata
\ion{O}{7} He$\alpha$/He$\beta$\tablenotemark{b}   & 6.54$\pm$0.76 & 8.54  & 6.52 &
5.22 & 2.49 \\
\ion{O}{8} Ly$\alpha$/Ly$\beta$\tablenotemark{b}    & 6.95$\pm$0.58 & 9.63  & 8.07 & 
6.82 & 4.55 \\
\ion{Ne}{9} He$\alpha$/He$\beta$\tablenotemark{b} & 7.38$\pm$0.35 & 11.14 & 10.07 &
9.23 & 7.34 \\
\ion{Ne}{10} Ly$\alpha$/Ly$\beta$\tablenotemark{b} & 8.24$\pm$0.52 & 12.93 & 11.89 &
10.91 & 8.93 \\
\ion{Ne}{10} Ly$\alpha$/Ly$\beta$\tablenotemark{c} & 8.24$\pm$0.52 & 7.061 & 6.49 &
5.96 & 4.88
\enddata
\tablenotetext{a}{Errors are 1$\sigma$.}
\tablenotetext{b}{Predictions assume $T_e = 2.5 \times 10^6$ K.}
\tablenotetext{c}{Predictions assume $T_e = 1.0 \times 10^7$ K.}
\end{deluxetable}

\begin{figure}
\epsscale{1.13}
\plotone{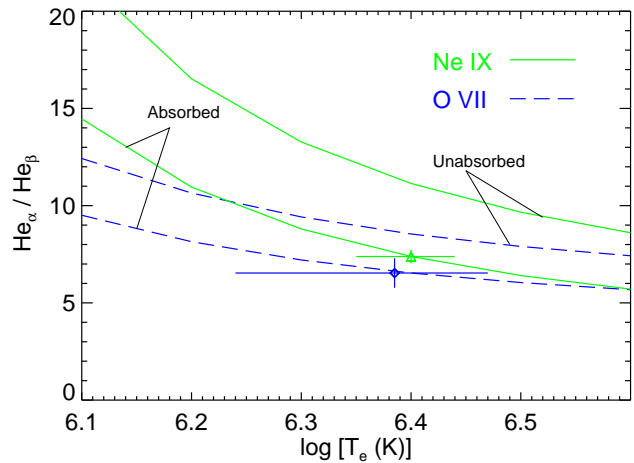}
\caption{Theoretical models of He$_\alpha$ to He$_\beta$ line flux
ratios as functions of $T_e$, with {\it solid} curve for \ion{Ne}{9}
and {\it dashed} curve for \ion{O}{7}.  The filled circle is the
observed ratio for \ion{Ne}{9} and the open circle is for \ion{O}{7},
with 1$\sigma$ error bars. Unabsorbed and best absorbed models for
each ion are labeled. The absorbed models shown have $N_H = 1.8 \times
10^{21}$ and $4.1 \times 10^{20}$ cm$^{-2}$ for \ion{Ne}{9} and
\ion{O}{7}, respectively.}
\end{figure}

Ratios of the H-like series lines (Ly$_{\alpha}$, Ly$_{\beta}$, etc.)
are also sensitive to both absorption and temperature, but unlike the
He-like ions, H-like ions do not present an independent temperature
diagnostic.  If the absorption column depends on the ion, as implied by the
inconsistency between \ion{Ne}{9} and \ion{O}{7}, then we need to
assume a model $T_e$ to obtain a column density from the H-like line
ratios.  Assuming 2.5~MK, the observed ratio \ion{O}{8}
Ly$_\alpha$/Ly$_\beta$ indicates $N_H = 7.8 ^{+1.1} _{-0.9} \times
10^{20}$ cm$^{-2}$. This value is consistent within the errors with
$N_H$ from \ion{O}{7} and thus a third absorber is not required. For \ion{Ne}{10} at 2.5~MK, the column density
would be larger than the column density for \ion{Ne}{9}, as shown in Table~4;
however, Table~4 also shows that no absorption is
required if the \ion{Ne}{10} lines are formed at 10~MK. We must further consider the effects of temperature on
\ion{Ne}{10} line ratios. 

We can also  estimate the temperature using the charge state
balance. Table~5 presents the He$_\alpha$/Ly$_\alpha$ ratios for
oxygen and neon. First, the observed ratios are corrected for different
column densities as found above, and then $T_e$ is derived from the
respective corrected ratios. For oxygen, good consistency is found for
the temperatures derived from the G-ratio and the \ion{O}{7}
He$_{\alpha}$/\ion{O}{8} Ly$_{\alpha}$ ratio, given a column density
of about 8 $\times 10^{20}$ cm$^{-2}$. For neon, on the other hand,
all the derived temperatures are larger than $T_e$ derived above from
the \ion{Ne}{9} G-ratio. Thus it appears that some of the \ion{Ne}{10}
emission originates at higher temperature.  Using $N_H = 2 \times
10^{21}$ cm$^{-2}$ and $T_e = 2.5$~MK, and the observed \ion{Ne}{9}
He$_{\alpha}$ flux, we predict the flux of the Ne~X Ly$_{\alpha}$ to
be only 18\% of the observed value. In this case, the rest of the
\ion{Ne}{10} emission must be formed at higher temperature. If the
trend of increasing $N_H$ with charge state 
continues, then $N_H$ for \ion{Ne}{10} would be higher than $2 \times
10^{21}$ cm$^{-2}$, and the
fraction of Ne~X formed at 2.5~MK would be somewhat larger. 

\begin{deluxetable}{lcccc}
\tabletypesize{\footnotesize}
\tablewidth{0pc}
\tablecaption{$T_e$ From Charge State Balance}
\tablehead{
Identification &
\colhead{Observed\tablenotemark{a}} &
\multicolumn{3}{c}{Corrected for log $N_H$ (cm$^{-2}$):} \\
\colhead{} &
\colhead{} &
\colhead{20.61 } &
\colhead{20.89} &
\colhead{21.26}
}
\startdata
\ion{O}{7} He$\alpha$/\ion{O}{8} Ly$\alpha$   & 0.55$\pm$0.052 &
0.70 & 0.85 & 1.67 \\
\ion{Ne}{9} He$\alpha$/\ion{Ne}{10} Ly$\alpha$  & 2.32$\pm$0.08 &
2.48 & 2.60 & 3.04 \\
\hline \hline \\
Identification & \multicolumn{4}{c}{Derived $T_e$ ($10^6$ K)\tablenotemark{b}} \\
\hline \\
\ion{O}{7} He$\alpha$/\ion{O}{8} Ly$\alpha$ & 3.03  & 2.83 & 2.66 & 2.23 \\
\ion{Ne}{9} He$\alpha$/\ion{Ne}{10} Ly$\alpha$  & 4.16 & 4.08 & 4.01 & 3.87
\enddata
\tablenotetext{a}{Errors are 1$\sigma$.}
\tablenotetext{b}{The derived $T_e$ values are based on the ratios
  listed above in the corresponding observed
  and corrected columns.}
\end{deluxetable}

The emissivity functions for lines of \ion{Fe}{17} and \ion{Ne}{10}
both peak near 5~MK, suggesting the possibility of comparing their
formation temperatures. The ratio of \ion{Fe}{17} $\lambda$15.014 to
$\lambda$15.261 is sensitive to $T_e$, due to a blend of an
inner shell excitation line of \ion{Fe}{16} at 15.261 \AA\
(Brown et al.\  2001). The temperature sensitivity essentially comes from
the charge state balance of \ion{Fe}{16} and \ion{Fe}{17}. Using the reported experimental dependence of
the line ratio on the electron beam energy of the Electron Beam Ion
Trap at Lawrence Livermore National Laboratory (Brown et al.\  2001), the observed HETG value gives $T_e
= 2.34^{+0.09}_{-0.06}$~MK. Despite similar peak temperatures,
\ion{Fe}{17} and \ion{Ne}{10} are not formed together. Apparently, the
long tail of the \ion{Ne}{10} Ly$_{\alpha}$
emissivity function toward high temperature contributes significantly
to its emission. The \ion{Fe}{17} lines
are predominantly formed in the accretion shock, while the \ion{Ne}{10} lines are
formed both in the shock and in the hotter corona.

One might expect that the ions formed at low temperature
(\ion{O}{7}, \ion{O}{8}, and \ion{Ne}{9}) should all experience similar absorption. Using $N_H$
derived from \ion{Ne}{9}, we can derive characteristic
emitting $T_e$ for \ion{O}{8} and \ion{O}{7} of 1.4~MK and $< 1.0$~MK,
respectively. These temperatures are far lower than $T_e$ derived from
the \ion{O}{7} G-ratios. They are both also far lower than $T_e > 2
$~MK derived from the charge state ratio of oxygen represented over
the range of $N_H$ found here, as shown in Table~5. We conclude that an
increase in $N_H$ with increasing charge state (presumably temperature) up through \ion{Ne}{9} is
consistent with all the data. 

We note that Argiroffi et al.\  (2009) claim evidence for resonance
scattering of the strongest lines in the XMM-RGS spectrum of MP~Mus,
though their RGS spectrum of TW~Hya is too weak to show the signatures
of absorption we find here. We rule out resonance scattering in TW~Hya
using the strong series of \ion{Ne}{9}. An optical depth $\tau$ of
about 0.4 for \ion{Ne}{9} He$_{\alpha}$ would be required to model the
observed He$_{\alpha}$/He$_{\beta}$ ratio. For a single ion, the
optical depth $\tau$ scales as $g f_{osc} \lambda$, such that we can
then predict the other \ion{Ne}{9} line ratios. We find then that
He$_{\alpha}$/He$_{\gamma}$ ratio is overpredicted by 50\%. Since the
oscillator strengths decrease rapidly with increasing principal
quantum number $n$, only the He$_{\alpha}$ line might be affected by
resonance scattering. We find instead, that all the \ion{Ne}{9} lines
are affected by absorption, and rule out resonance scattering as the
absorption mechanism. We conclude that the absorber is neutral or near
neutral, consistent with the preshock accreting gas, as suggested by
theoretical studies. For example, Lamzin (1999) considers the
absorption by the preshock gas of the X-ray emitting plasma for
different geometries and orientations. Gregory et al.\  (2007) use
realistic coronal magnetic fields coupled with a radiative transfer
code to calculate the obscuration of X-ray emission by accretion
columns, and suggest that this effect can explain the observed low
X-ray luminosities of accreting young stars relative to non-accretors.

\subsection{Emission measure (EM)  distribution and elemental
abundances}

We present here four models for the emission measure (EM) distribution\footnote{Emission measure EM
is defined as $\int{0.8 N_e^2 dV}$, where the factor 0.8 accounts for
the hydgrogen to electron density ratio and the integral is taken over
the volume $V$. The intensity of a spectral feature $I = \varepsilon$
EM $/ (4 \pi D^2)$ where $\varepsilon$ is the emissivity in units of
ph cm$^3$ s$^{-1}$ and $D$ is the distance to the source. The function
$\varepsilon$ depends on $T_e$, and is given in steps of log [$T_e$
(K)] = 0.1 in ATOMDB (Smith et al.\  2001).}  and elemental abundances (Table~6). These
models serve to illustrate how values derived from the line ratios
affect the global fit to the spectrum and to show whether abundance
determinations are robust. For all four models, the first-order HEG
and MEG spectra are fit to a set of variable abundance APEC models,
using \Sherpa\ and applying different constraints to the fitted
parameters. All models have acceptable goodness-of-fit values. Model A
has two $T_e$ components and a single absorber.  Model~A should be
appropriate for comparison with other X-ray spectra of cool stars obtained at
lower spectral resolution.

\begin{deluxetable*}{lccccc}
\tabletypesize{\footnotesize}
\tablewidth{0pc}
\tablecaption{Model Parameters\tablenotemark{a} from Global Spectral Fitting}
\tablehead{
\colhead{Parameter} &
\colhead{Model A} &
\colhead{Model B} &
\colhead{Model C} &
\colhead{Model D}
}
\startdata
$T$1 (MK)                    & 3.58$\pm$0.04   & 2.5\tablenotemark{b} & 2.5\tablenotemark{b} & 2.5\tablenotemark{b} \\
EM1    ($10^{53}$~cm$^{-3}$)   & 1.11$\pm$0.13   & 1.49$\pm$0.09        & 1.56$\pm$0.10  &   4.22$\pm$0.70    \\
$T$2 (MK)                    & 18.8$\pm$0.5    & 11.2$\pm$0.2         & 12.6\tablenotemark{c} & 12.6\tablenotemark{d} \\
EM2    ($10^{53}$~cm$^{-3}$)   & 0.33$\pm$0.01 & 0.47$\pm$0.01          & 0.121$\pm$0.004\tablenotemark{c} & 0.101$\pm$0.005\tablenotemark{d} \\
$N_H$  ($10^{21}$~cm$^{-2}$)   & 0.15$\pm$0.22   & 1.0\tablenotemark{b} & 1.0\tablenotemark{b} & 1.0\tablenotemark{b} \\
N  (8.05)\tablenotemark{e}            & 0.66$\pm$0.14   & 0.28$\pm$0.06        & 0.53$\pm$0.10 & 0.20$\pm$0.04 \\
O  (8.93)\tablenotemark{e}            & 0.23$\pm$0.02   & 0.14$\pm$0.01        & 0.24$\pm$0.02 & 0.09$\pm$0.01\\
Ne (8.09)\tablenotemark{e}           & 1.23$\pm$0.08   & 2.04$\pm$0.11        & 2.50$\pm$0.13 & 1.04$\pm$0.02\\
Mg (7.58)\tablenotemark{e}           & 0.18$\pm$0.02   & 0.16$\pm$0.02        & 0.21$\pm$0.02 & 0.18$\pm$0.02\\
Si (7.55)\tablenotemark{e}           & 0.33$\pm$0.03   & 0.24$\pm$0.02        & 0.29$\pm$0.03 & 0.34$\pm$0.03\\
Fe (7.67)\tablenotemark{e}           & 0.13$\pm$0.01   & 0.10$\pm$0.01        & 0.20$\pm$0.01 & 0.16$\pm$0.01
\enddata
\tablenotetext{a}{All errors given are statistical errors from the
  fit.}
\tablenotetext{b}{Value is fixed.}
\tablenotetext{c}{EM2 given is the fit to the peak of the EM distribution. Shape of the hot component is the same as for Model D, as described in the text.}
\tablenotetext{d}{EM2 given is the fit to the peak of the EM
  distribution. Shape of the hot component EM distribution is shown in Figure~7 and described in the text.}
\tablenotetext{e}{Abundance of element relative to the solar values of
    Anders \& Grevesse (1989). The abundances in logarithmic form
    with H=12.0 are given in parenthesis next to the element label.}
\end{deluxetable*}

Model~B is also a two-$T_e$ model, but with constraints imposed from
the line ratios. The lower $T_e$ is fixed at 2.5~MK, and the single
absorber $N_H$ is fixed at 1.0 $\times 10^{21}$~cm$^{-2}$ (a rough
average of the values obtained from line ratios). The $N_e$-sensitive
forbidden and intercombination line regions are excluded from the fit.

Table~6 also gives results from multicomponent Models~C and D, each
with the low $T_e$ and a single $N_H$ fixed, as in Model~B.  Assuming
that the hotter component is coronal in nature, a broad distribution
of $T_e$ is reasonable, although the shape is not well constrained. We
broaden the hot component using a Gaussian-shaped function centered
around 12.5~MK. For Model~C the entire spectrum is fit simultaneously,
with only the forbidden and intercombination line regions excluded as
for Model~B.

\begin{figure}
\epsscale{1.13}
\plotone{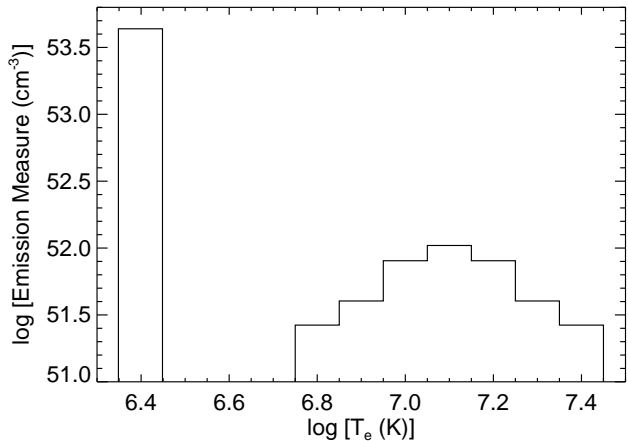}
\caption{EM distribution for Model~D as described in text.}
\end{figure}

For Model~D the emission measures for the two components are fit to
line-free regions, identified visually and using APEC models. The
thermal continuum emission fit to line-free regions is strongly
temperature-dependent and thus constrains the emission measure
distribution.  The abundances are then fit to narrow regions centered
on the strong lines. These separate fitting procedures are iterated
until the values stop changing.  The resulting Model~D EM distribution
is shown in Figure~7. The larger statistical error for the low $T_e$
emission measure in Model D compared with Model~C (Table~6) stems from
the smaller number of bins used in the fits; however, we expect the
systematic errors for Model~D to be lower, given that we are only
using bins whose information content is secure, and thus we choose
Model~D to illustrate the model spectrum.\footnote{The four models are
all statistically acceptable; however, it is interesting to note that
none of the models do a good job of fitting all the emission line
fluxes. For example, Model~D does underpredicts \ion{Ne}{10}
Ly$_\alpha$ by almost a factor of 2. We attribute these difficulties
primarily to the complexity of the absorption, and secondarily to the
few constraints on the shape of the EM distribution.} Figure~8
compares the Model D spectra predicted by the broad ``coronal''
component and the soft ``accretion'' component with the observed
spectra. Table~1 lists the origin of each line as accretion or corona
based on Model D.  An additional component with $T_e$ lower than
2.5~MK can be added to both Models C and D but is not
required, as it does not improve the fit significantly. Similar results are obtained with somewhat different choices
of width and peak $T_e$ for the hot component.

\begin{figure}
\epsscale{1.12}
\plotone{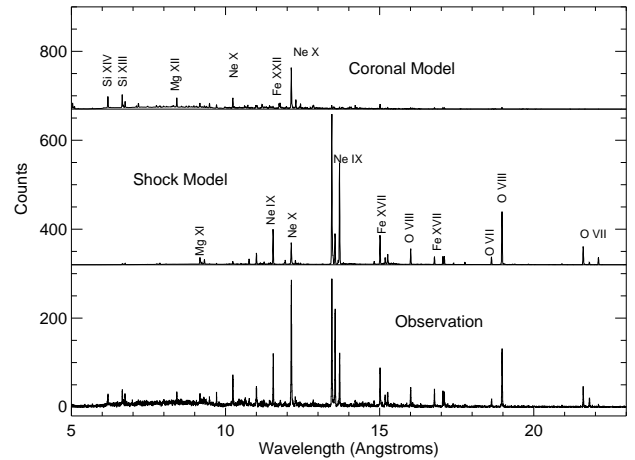}
\caption{Comparison of predicted and observed spectrum from the Medium
Energy Grating (MEG). Upper curve is the predicted spectrum from the
Model~D EM distribution above 5~MK (the ``coronal model''), shown in
Figure~7. Middle curve is the predicted spectrum from the Model~D
emission measure at 2.5~MK (the ``shock model''). The lower curve is
the observed spectrum. As noted in the text, Model~D significantly underpredicts the
\ion{Ne}{10} Ly$_\alpha$. Also note that the forbidden and
intercombination line predictions using the  APEC models in \Sherpa\
are at their low density values (e.g., \ion{Ne}{9}).}
\end{figure}

Since the line ratios indicate increasing $N_H$ with higher charge
state, we tried adding a second absorber to the high $T_e$ Gaussian
component for Models B, C, and D. Robust solutions are difficult to
obtain, as the absorption, abundances, and normalizations of the
emission measures do not independently determine the global X-ray
spectrum. We also explored the possibility that the low and high $T_e$
components have different metal abundances but again, without robust
results.

Using $T_e$=2.5~MK, as derived from the \ion{Ne}{9} line ratio,
instead of $T_e$=3.58~MK as found in Model~A, has important
implications for interpreting the formation of some of the emission
lines.  The ratio of the strong \ion{Ne}{9} He$_{\alpha}$ and
\ion{Ne}{10} Ly$_{\alpha}$ line emission drives the Model~A fit to
more than 3.0~MK with very little absorption. Specifically, only
$\sim$15\% of the \ion{Ne}{10} Ly$_\alpha$ emission arises from the
2.5~MK component, whereas a 3.58~MK component can produce more than
half of the \ion{Ne}{10} emission. The 2.5~MK component produces
essentially all of the \ion{N}{7}, \ion{O}{7}, \ion{O}{8}, and
\ion{Ne}{9} emission and more than half the emission from
\ion{Fe}{17}. The fluxes from these lines formed at 2.5~MK show no
increase over their average value during the flare, as shown in
Figure~1, indicating that
the flare is associated with the hotter corona rather than the
accretion shock, and justifying its retention in our analysis. Lines
from \ion{Mg}{12}, \ion{Si}{13}, \ion{Si}{14} (Fig.~1), and \ion{Fe}{22} are produced by the hotter
component and may participate in the flare. While K02 found that \ion{Mg}{11} was overpredicted by
their model by a factor of about 3, our Models B, C, and D have no
such problem. In fact, for Models C and D, the small flux of the
\ion{Mg}{11} resonance line forces the emission measure to be
negligible between $\sim$3 and 5~MK (Fig.~7).

The relative elemental abundances given in Table~6 are reasonably
consistent for the different models. The abundances are also similar
to those found by K02, Stelzer \& Schmitt (2004), and Drake et
al. (2005).  It is important to note that the continuum emission
arises primarily from the hot component, except at the longest
wavelengths (Fig.~8). Thus absolute abundances in the low $T_e$
component for Ne, O, Fe and other species cannot be determined
reliably. Nevertheless, the extremely large Ne/O abundance ratio
appears to be a very robust result. Differences in the absolute
abundances derived here reflect both differences in $T_e$ and the
degeneracy between emission measure and $N_H$. For example, nitrogen
and oxygen are formed entirely in the low temperature component and
thus the abundance differences among the models reflect differences
among the low temperature emission measure and $N_H$, as does the
difference in the neon abundance between Models A and B. The
difference in the neon abundance between Models C and D also reflects
the differences in their emission measures at 2.5~MK.

Weak emission is apparent from the S and Ar complexes, but we have not
included line fluxes from these elements in Table~1 because the
individual lines (in particular the He-like resonance line) cannot not
be cleanly determined in the four grating arms. Instead we have used
the Model C global fit to obtain 0.21 $\pm$ 0.09 and 1.41 $\pm$ 0.49
times solar (Anders \& Grevesse 1989) for the S and Ar abundances,
respectively. A
high Ar/O abundance would add weight to Drake et al.'s (2005)
suggestion that the accretion stream is depleted of grain-forming
elements; however, the high Ar abundance is unfortunately not
statistically significant.

\subsection{Velocity constraints}

In this section, we discuss velocity constraints from the line
measurements. The measurements of the emission line fluxes given in
Table~1 use Gaussian lines in addition to the standard HETG
calibration line spread function. Table~7 gives the lines with the
highest signal-to-noise ratios, for which Gaussian widths are
measured. Lines not listed do not show statistically significant
widths, but are consistent with the range of widths found here. For
these fits the continuum level is allowed to vary to provide a better
local fit. Assuming the \ion{Ne}{9} lines form in the same
region, and using a flat continuum flux level from 13.45 to 13.75 \AA,
we force the 3 triplet lines to have the same Gaussian width to
improve the significance of the velocity measurement. Figure~2 shows
the best fit for the \ion{Ne}{9} lines. As before, the centroids of
the lines measured in each grating arm are fit independently. We thus
obtain a velocity of 183 $\pm$ 16 km s$^{-1}$. The thermal Doppler
FWHM velocity for neon at 2.5~MK is 75 km s$^{-1}$. The error in the
calibration is estimated to be about 5\%. Thus we determine a
turbulent velocity of 165 $\pm$ 18 km s$^{-1}$, including calibration
error in quadrature with the 1$\sigma$ statistical error. The line
profiles are consistent with each other within the errors and suggest a range of velocities from zero up to
about the sound speed of the gas (185 km s$^{-1}$ at 2.5~MK). The turbulent velocity is well below
the preshock gas velocity of $\sim$500 km s$^{-1}$. Line centroids are consistent with
reference wavelengths from ATOMDB to within the calibration
uncertainty on the absolute wavelength scale.

\begin{deluxetable*}{lrcccc}
\tabletypesize{\footnotesize}
\tablewidth{0pc}
\tablecaption{Velocities from Strong Emission Lines}
\tablehead{
\colhead{Line} &
\colhead{$\lambda_{ref}$} &
\colhead{FWHM} &
\colhead{V$_{obs}$\tablenotemark{a}} &
\colhead{V$_{th}$\tablenotemark{b}} &
\colhead{V$_{turb}$\tablenotemark{c}} \\
\colhead{} &
\colhead{(\AA)} &
\colhead{(\AA)} &
\colhead{(km s$^{-1}$)} &
\colhead{(km s$^{-1}$)} &
\colhead{(km s$^{-1}$)} \\
}
\startdata
\ion{Ne}{10} & 10.239    & 0.0042 $\pm$ 0.0030 & 122 $\pm$ 87 & 75\tablenotemark{d} & 96 \\
\ion{Ne}{ 9} & 11.544    & 0.0024 $\pm$ 0.0024 &  62 $\pm$ 62 & 75 & 0 \\
\ion{Ne}{10} & 12.134    & 0.0082 $\pm$ 0.0009 & 203 $\pm$ 23 & 75\tablenotemark{d} & 189 \\
\ion{Ne}{ 9} & 13.447    & 0.0103 $\pm$ 0.0010 & 229 $\pm$ 22 & 75 & 216 \\
\ion{Ne}{ 9} & 13.553    & 0.0057 $\pm$ 0.0014 & 126 $\pm$ 32 & 75 & 101 \\
\ion{Ne}{ 9} & 13.699    & 0.0073 $\pm$ 0.0019 & 159 $\pm$ 41 & 75 & 140 \\
\ion{Fe}{17} & 15.014    & 0.0065 $\pm$ 0.0040 & 129 $\pm$ 80 & 45 & 121 \\
\ion{O }{ 8} & 18.969    & 0.0074 $\pm$ 0.0026 & 116 $\pm$ 41 & 85 & 79 \\
\enddata
\tablenotetext{a}{V$_{obs}$ is the observed FWHM velocity.}
\tablenotetext{b}{V$_{th}$ is the thermal FWHM velocity calculated at 2.5~MK.}
\tablenotetext{c}{V$_{turb}$ is the turbulent FWHM velocity assuming $V_{turb}=\sqrt{(V_{obs}^2-V_{th}^2)}$.}
\tablenotetext{d}{Note the thermal velocity for neon at 10~MK is 150 km s$^{-1}$.}
\end{deluxetable*}

\subsection{Physical conditions of the X-ray plasma}

We summarize here the physical conditions of the X-ray plasma from a
consistent analysis of the spectra. The new diagnostic measurements
strongly support K02's argument that the low temperature X-ray component
($T_e \sim$ 2.5~MK) arises in the accretion shock. While the density we derive from
\ion{Ne}{9}, $N_e = 3 \times 10^{12}$ cm$^{-3}$, is within the range
found in active stars at higher $T_e$ ($\sim$ 6~MK), it is at least an order of
magnitude larger than the few \ion{Ne}{9} R-ratios reported
(Huenemoerder et al.\  2001; Ness et al.\  2002; Osten et al.\  2003). The
$N_e$ derived from the \ion{O}{7} R-ratio is also more than an order
of magnitude larger than derived from \ion{O}{7} in other cool stars (see 
Sanz-Forcada et al.\  2003; Ness et al.\  2004; Testa et al.\  2004). We also measure the \ion{Mg}{11}
R-ratio and rule out photoexcitation, since the observed \ion{Mg}{11}
R-ratio can only be produced by high $N_e$, namely $5.8 \times
10^{12}$ cm$^{-3}$.

The G-ratios from \ion{O}{7}, \ion{Ne}{9}, and \ion{Mg}{11} give
similar $T_e \sim 2.5$~MK, significantly lower than the $T_e$ of peak
emissivity for \ion{Ne}{9} (4~MK) and \ion{Mg}{11} (6.3~MK). The
G-ratio from \ion{Ne}{9} is particularly reliable because of good
statistics and accurate atomic data. In our multi-$T_e$ models (Models
C and D), this low $T_e$ component is isolated and does not connect
continuously with the hotter, coronal component. Active main-sequence
stars tend to have emission measures rising up to 6~MK and above, such
that, if biased at all, $T_e$-sensitive line ratios should be biased
toward higher rather than lower temperature. Thus the $T_e$ measurements
also support the accretion shock scenario. More accurate G-ratio
models for \ion{O}{7} and \ion{Mg}{11} are required to establish any
$T_e$ differences among these ions.

We also measure strong absorption using line ratios from several ions,
and require at least two different absorbing column densities, with the higher
charge states experiencing more absorption. It seems likely that the
gap between the two components of the EM distribution is caused by
neutral H absorption of
the soft X-ray spectrum that is produced by the hot coronal component.  The
pattern of absorption is reminiscent of the ``two-absorber X-ray
sources'' reported by G\"{u}del et al.\  (2007b), where the coronal
component has ten times more absorption than the soft component;
however, unlike TW~Hya, these sources have high accretion rates and
are believed to drive jets. We note that the absorption in TW~Hya is
identified here from line ratio diagnostics, whereas the 
absorption in G\"{u}del et al.\  (2007b) was determined from CCD spectra. We were not able to
determine multiple absorbers using standard global fitting methods,
even with the high resolution spectrum presented here, because the
continuum spectrum has low signal-to-noise ratio. On the other hand, it
seems likely that with low spectral resolution, multiple absorbers
could also be easily missed.  Thus the presence of more than one
absorber may be a more universal characteristic than we can presently
establish without reliable line ratio diagnostics.

All the values of $N_H$ found here are higher than those found using
the hydrogen Ly$\alpha$ profile, for which a conservative upper limit
is $6 \times 10^{19}$ cm$^{-2}$ (Herczeg et al.\  2004), indicating that
the absorption is not due to the interstellar medium but is intrinsic
to the stellar system. Our column densities are consistent
with the range of values previously reported from ROSAT and ASCA
(Kastner et al.\  1999).  In fact the two different values from ROSAT
and ASCA are consistent with our different values and may reflect the
complex absorption coupled with the different instrument responses rather than time-variable absorption. As noted also
by G\"{u}del et al.\  (2007b) for the highly absorbed coronae of DG Tau
A, GV Tau, DP Tau, our X-ray
measurements are higher than expected from the optical extinction
assuming standard gas-to-dust ratios. For X-ray emission that is
highly localized compared with the stellar Ly$_\alpha$, and
preferentially absorbed by the accretion stream directly above it, a
difference in $N_H$ derived by the two methods seems entirely
reasonable.

\section{The Accretion Shock Model}

We can test the hypothesis that some of the measured X-ray emission
comes from the accretion shock by constructing a standard
one-dimensional (1D) model of the magnetospheric infall, shock
heating, and postshock cooling.  This model is a simplified version of
similar 1D models in the literature (Calvet \& Gullbring 1998;
G\"{u}nther et al.\ 2007), and is tailored to specific measured
properties of TW Hya.

The magnetospheric accretion is assumed to follow a set of dipole
magnetic field lines that thread the accretion disk.  We use
K\"{o}nigl's (1991) expression for the inner truncation radius of the
disk to determine the distance at which parcels of gas are launched
from rest.  This expression depends on the mass accretion rate
$\dot{M}_{\rm acc}$, the surface magnetic field strength $B_{\ast}$,
and the stellar mass $M_{\ast}$ and radius $R_{\ast}$.  For TW Hya, we
take $M_{\ast} = 0.7 \, M_{\odot}$ and $R_{\ast} = 0.8 \, R_{\odot}$
(Batalha et al.\ 2002).  As in Cranmer (2008), we assume a canonical T
Tauri star magnetic field strength of $B_{\ast} \approx 1000$ G.  A
range of accretion rates for TW Hya has been reported, from $4 \times
10^{-10} \, M_{\odot}$ yr$^{-1}$ (Muzerolle et al.\ 2000) to $5 \times
10^{-9} \, M_{\odot}$ yr$^{-1}$ (Batalha et al.\ 2002).  The low end
of this range produces the best agreement with the observations
presented here, so the Muzerolle et al.\ accretion rate will be
adopted for the cooling-zone models below.

With the above values, the K\"{o}nigl (1991) truncation radius of the
disk is
found to be approximately $r_{t} = 4.5 \, R_{\ast}$, which implies
a ballistic free-fall velocity at the stellar surface of
\begin{equation}
  v_{\rm ff} = \left[ \frac{2 G M_{\ast}}{R_{\ast}}
  \left( 1 - \frac{R_{\ast}}{r_t} \right) \right]^{1/2}
  \, \approx 509 \,\, \mbox{km} \,\, \mbox{s}^{-1}  \,\, .
\end{equation}
Since the preshock gas is often assumed to have a sound speed only
of order $\sim 10$ km s$^{-1}$ (e.g., Muzerolle et al.\  2001),
the highly supersonic flow should produce a strong shock at the
stellar surface, heating the plasma to a postshock temperature of
$T_{\rm post} \approx 3 m v_{\rm ff}^{2} / 16k$,
where $k$ is the Boltzmann constant and $m$ is the mean atomic
mass.  For the above parameters, this value is $T_{post} =$ 3.4~MK.
[We neglect any preshock heating due to photoionization since the
X-ray luminosity is relatively low. See Calvet \& Gullbring (1998).]

The preshock mass density $\rho_{\rm pre}$ in the accretion stream can be
estimated via mass flux conservation at the surface, i.e.,
\begin{equation}
  \dot{M}_{\rm acc} = 4\pi f R_{\ast}^{2} \rho_{\rm pre} v_{\rm ff}
\end{equation}
where the $f$ is the filling factor of the stellar photosphere heated
by the accretion stream. The value of  $f$ was estimated to be 1.1\%
for the dipole magnetic geometry assumed above (see also Cranmer 2008).
Assuming complete ionization (with 10\% He by number), the
preshock electron number density $N_{\rm pre} = 5.7 \times 10^{11}$
cm$^{-3}$.
Because the shock is strong, we obtain an immediate postshock density
$N_{\rm post} = 2.3 \times 10^{12}$ cm$^{-3}$ (i.e., a factor of
four increase across the shock, with the assumed adiabatic index of
$5/3$).  The postshock velocity $v_{\rm post}$ is thus approximately
130 km s$^{-1}$.

As one proceeds deeper through the cooling zone, the density increases
and the velocity and temperature decrease, until a depth is reached at
which there is no longer any influence from the shock.
The ``bottom'' density $\rho_{\rm bot}$ can be estimated using
ram pressure balance between the accretion stream and the unperturbed
atmosphere. Using Equation~2 the pressure of the accretion stream is given by 
\begin{equation}
  P_{\rm ram} \, = \, \frac{\rho_{\rm pre} v_{\rm ff}^2}{2} \, = \,
  \frac{v_{\rm ff} \dot{M}_{\rm acc}}{8 \pi f R_{\ast}^2}
\end{equation}
(e.g., Hartmann et al.\  1997; Calvet \& Gullbring 1998).
Assuming the accretion is stopped in the first few scale heights
above the photosphere (at which radiative equilibrium drives the
temperature to $\sim (4/5) T_{\rm eff}$), the pressure of the atmosphere
is $0.8 \rho_{\rm bot} k T_{\rm eff}$. By equating the two pressures, the density can be found
by solving for 
\begin{equation}
  \rho_{\rm bot} \approx \frac{ P_{\rm ram}}{0.8 k T_{\rm eff}}
  \label{eq:rhosh}
\end{equation}
with $T_{\rm eff} = 4070$ K for TW Hya (Batalha et al.\  2002;
see also Cranmer 2008).  Converting to electron density, this
corresponds to  $2.6 \times 10^{15}$ cm$^{-3}$, a factor of 1000 higher
than the immediate postshock density.

Finally, we computed the physical depth and structure of the postshock
cooling zone using the analytic model of Feldmeier et al.\ (1997),
which assumes that the radiative cooling rate $Q_{\rm rad} \approx N_{e}^2
\Lambda(T)$ has a power-law temperature dependence ($\Lambda
\propto T^{-1/2}$).  The thickness of the cooling zone is proportional
to $v_{\rm post}^{4} / \rho_{\rm post}$, and for the case of the
TW~Hya  parameters used here, it has a value of 385 km.  The spatial
dependence of density, temperature, and velocity in the cooling zone
is specified by equations (8)--(11) of Feldmeier et al.\ (1997), and
the shapes of these curves resemble those of G\"{u}nther et al.\
(2007)---which were computed using a more detailed radiative cooling
rate---reasonably well (see Fig.~9).

Note that choosing the stronger mass accretion rate of Batalha et al.\
(2002) would have led to a truncation radius closer to the star
($r_{t} = 2.2 \, R_{\ast}$), a lower freefall velocity ($v_{\rm ff} =
430$ km s$^{-1}$), and thus a lower postshock temperature ($2.4 \times
10^{6}$ K).  The densities would all have been higher than their
counterparts above by about an order of magnitude, and the thickness
of the cooling zone would have been smaller (32 km).

Using the temperature and density in the model, we compute the line
intensities as functions of distance from the accretion shock. Figure~9 shows line intensities
through the region for the strongest lines of
\ion{Ne}{9}, \ion{O}{8}, and \ion{O}{7}. We compare the integrated
intensities to the observations. 
Assuming the abundances determined from Model~D, and using $N_H = 1.0
\times 10^{21}$ cm$^{-2}$, the
predicted line intensities for the \ion{Ne}{9}, \ion{O}{8} and \ion{O}{7}
resonance lines are 0.25, 0.16, and 0.40 times the observed fluxes,
respectively. Higher abundances improve the agreement, as do lower
column densities. Nevertheless, agreement within a factor of 4 seems
quite reasonable, given the difficulty of obtaining absolute
abundances and the complexity of the absorption.

\begin{figure}
\epsscale{1.15}
\plotone{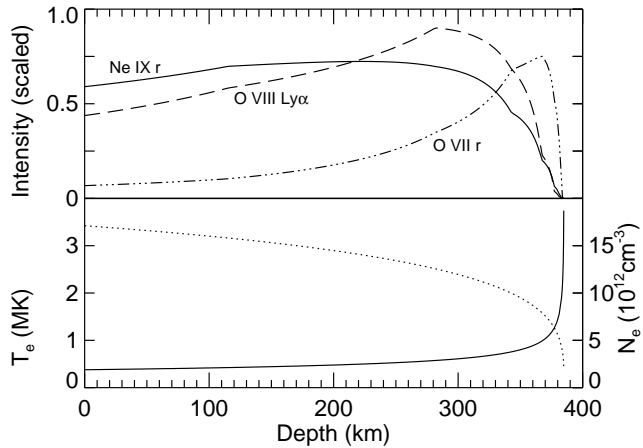}
\caption{{\it Lower:} $T_e$ (dotted) and $N_e$ (solid) as functions of
depth in the postshock cooling column model, where 0 corresponds to
the position of the shock. The model is described in Section 4.
{\it Upper:} Fractional contributions to the line intensities as
labeled, as functions of depth in the model.}
\end{figure}

The predicted G-ratios give $T_e =$ 2.5 and 2.2~MK for \ion{Ne}{9} and
\ion{O}{7}, respectively.  The $N_e$ calculated for \ion{Ne}{9}
is $2.6 \times 10^{12}$ cm$^{-3}$, in excellent agreement with the
observation; however, for \ion{O}{7}, the calculated $N_e$ is $4.1
\times 10^{12}$ cm$^{-3}$, seven times larger than observed. The
\ion{O}{7} density is clearly discrepant with the standard model.

\section{Analysis of the Postshock Cooling Plasma}

The physical conditions predicted at the shock front are in excellent
agreement with the observations; however, the predictions for the
postshock cooling plasma are in stark disagreement with the
observations. Most problematic is that the electron density derived
from \ion{O}{7} is {\it lower} than that for \ion{Ne}{9}, when the
postshock gas should be compressing as it cools and slows and hence
the density from \ion{O}{7} should be larger than from \ion{Ne}{9}.  Furthermore, $N_H$
determined from \ion{O}{7} is four times {\it smaller} than that
determined from \ion{Ne}{9}, whereas in the standard postshock model
\ion{O}{7} should be observed through the same or larger column
density of intervening material.  In this section, we explore the
properties of the postshock cooling plasma based on our new results.

Using the APEC code to calculate emissivities for a relevant grid of
$T_e$ and $N_e$, we construct a set of two-component shock models for
comparison with the diagnostic lines from the shock. The two
components represent Region~1, the higher $T_e$ region near the shock
front, and Region~2, the lower $T_e$ region of
postshock plasma. We do not discuss the hot coronal plasma here,
but  consider the  diagnostic lines from ions that are formed below 3~MK, in
particular \ion{O}{7}, \ion{O}{8}, \ion{Ne}{9}, and \ion{Mg}{11}. Each
component in the model has four
parameters: $T_e$, $N_e$, $N_H$, and volume $V$. The two regions
experience different absorption with a factor of 2.5 larger $N_H$ at
higher $T_e$, as required by the resonance series lines. The
abundances are taken from the Model D fits. The predicted fluxes from
the two regions are summed for comparison with the
observations. Table~8 gives the parameters for the best-fitting model,
the predicted fluxes for each of the two regions, and the ratios of
predicted to observed line fluxes. For \ion{O}{8}, \ion{Ne}{9}, and 
\ion{Mg}{11} the predictions agree to better than 30\% with 
the observed line fluxes. For \ion{O}{7} the predictions are low by a factor 
of less than 2, but the line ratios agree within 30\%.
The diagnostic line fluxes place tight constraints on the model.

\begin{deluxetable}{lcccc}
\tabletypesize{\footnotesize}
\tablewidth{0pc}
\tablecaption{Two-Region Model for Accretion Shock and Postshock Cooling}
\tablehead{
\multicolumn{5}{c}{Model Parameters\tablenotemark{a}} \\
\colhead{} &
\colhead{$N_e$} &
\colhead{$T_e$} &
\colhead{V} &
\colhead{$N_H$} \\
\colhead{} &
\colhead{(cm$^{-3}$)} &
\colhead{(MK)} &
\colhead{(cm$^3$)} &
\colhead{(cm$^{-2}$)}  
} 
\startdata
Region 1 & $6.0 \times 10^{12}$   & 3.00  & $1.5  \times 10^{28}$   &
$1.0 \times 10^{21}$  \\
Region 2 & $2.0 \times 10^{11}$   & 1.75  & $4.5  \times 10^{30}$   &
$4.0 \times 10^{20}$  \\
\hline \hline 
\multicolumn{5}{c}{Predicted Unabsorbed Fluxes\tablenotemark{b} from Region 1} \\
 & r & i & f & Ly$\alpha$ \\
O   & 260. & 166. & 3.6 & 512.\\ 
Ne  & 447. & 260. & 73.9 & \nodata \\
Mg  & 3.6  & 1.3  & 1.8 & \nodata \\
\hline \hline
\multicolumn{5}{c}{Predicted Unabsorbed Fluxes\tablenotemark{b} from Region 2} \\
 & r & i & f & Ly$\alpha$ \\
O &  124. & 70.8 & 36.7 & 25.1\\ 
Ne  &16.1 & 4.8& 10.1&  \nodata\\
Mg &0.0  & 0.0 & 0.0 & \nodata \\
\hline \hline
\multicolumn{5}{c}{Ratio of Predicted Absorbed Fluxes\tablenotemark{b} to Observed Fluxes\tablenotemark{b}} \\
 & r & i & f & Ly$\alpha$ \\
O & 1.00 & 0.69  & 0.53 & 0.69  \\ 
Ne  & 1.31 & 0.85\tablenotemark{c} & 0.75 \tablenotemark{c} & \nodata \\
Mg & 1.28 & 1.14 & 1.13 & \nodata 
\enddata
\tablenotetext{a}{Abundances from Model D are 0.09, 1.04, and 0.18 relative to
  solar (Anders \& Grevesse 1989), for oxygen, neon, and magnesium, respectively.}
\tablenotetext{b}{Fluxes at Earth are in units of $10^{-6}$ ph cm$^{-2}$ s$^{-1}$.}
\tablenotetext{c}{Chen et al.\  (2006) is about
  25\% larger than APEC.}
\end{deluxetable}

In this simple model, all of the emission from \ion{Mg}{11} and most
of the emission from \ion{Ne}{9} come from Region 1, which we
associate with the shock front. If we assume a cooling length of
200~km, based on the accretion shock model described above (see
Fig.~9), and use the derived emitting volume of $1.5 \times 10^{28}$ cm$^{3}$, we
obtain an area filling factor for Region 1 of 1.5\% of the surface area
of the star.

Region~2 has a larger volume than Region~1 by a factor of 300 as
determined primarily from \ion{O}{7}. If we
assume that the cooling zone is the continuation of Region~1, we again
obtain a length of about 200~km; however, this short length implies a
filling factor larger than the stellar surface area and is not
plausible. Instead, we use the scale height of 0.1 $R_\star$ for 1.75~MK
plasma to estimate a filling factor of 6.8\% for Region 2, a value in
good agreement with estimates from the ultraviolet (Costa et al.\  2000)
and optical continuum veiling (Batalha et al.\  2002).
Scaling the mass as $N_e V$ implies that the mass of Region~2 is 30
times the mass of Region~1. 

Romanova et al.\  (2004) have used
three-dimensional magnetohydrodynamic (MHD) simulations to show that
the hot spots formed on the surface by accretion are inhomogeneous,
with the lower density and temperature regions filling a larger
surface area than the denser central region, seemingly suggestive of Region
2. However, the larger mass of Region 2 also implies a higher
accretion rate than for Region 1, and thus a larger density, in contradiction to our
diagnostics. Therefore, we do not interpret Region~2 as a second
accretion shock with a different accretion rate, but rather as a new
accretion-driven phenomena. The impact from the accretion
presumably provides the energy to heat additional material in
the stellar atmosphere to over 1~MK. Very recently, two-dimensional
MHD simulations have shown that violent outflows of shock-heated
material can propagate from the base of the accretion shock
for cases with a high thermal to magnetic pressure ratio $\beta$
(Orlando et al.\  2009).

One of the key concepts of the standard one-dimensional accretion model suggests a
simple picture (Fig.~10). The field line that truncates the disk
separates the open magnetic field lines of the polar region from the
closed magnetic field loops in the equatorial regions. The
region heated by the shock (our Region 2) provides a large supply of
ionized material available to form the soft X-ray emitting corona
presented here, a warm wind (Dupree et al.\  2005), and soft X-ray jets
(G\"{u}del et al.\  2008). This corona exists because magnetic loops
confine it, but it is fed by the accretion process and is thus unlike
stellar coronae in dwarf stars on the main sequence. Figure~10 illustrates the
shock and its surroundings.

\begin{figure}
\epsscale{1.13}
\plotone{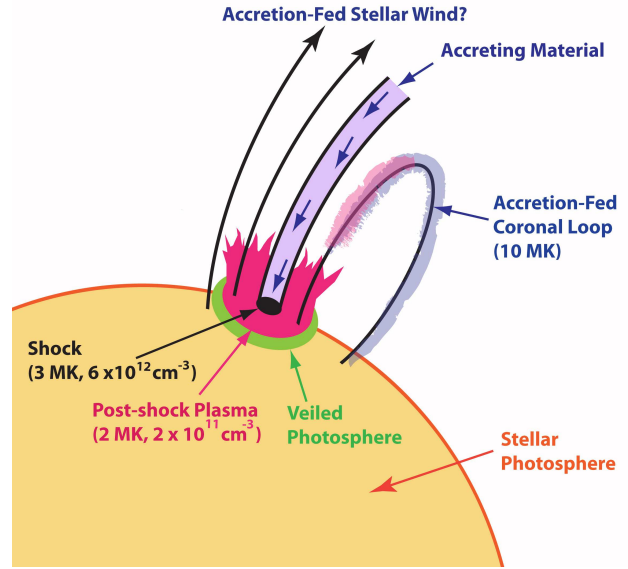}
\caption{Illustration of the shock and its surrounding
environment. Region~1 in the text corresponds to the shock and
Region~2 to the postshock plasma. Figure courtesy of A.\  Szentgyorgyi.}
\end{figure}

This model provides an explanation for the soft X-ray excess
discovered in the \xmm\ Extended Survey of the Taurus Molecular Cloud
(XEST; G\"{u}del et al.\  2007c; Telleschi et al.\  2007).  Grating spectra show that this excess emission
manifests itself in CTTS as enhanced \ion{O}{7} relative to \ion{O}{8}
as compared with weak-line T~Tauri (WTTS) and main-sequence stars (G\"{u}del \&
Telleschi 2007; Robrade \& Schmitt 2007).  Our
\ion{O}{7} measurements also demonstrate that the postshock cooling
gas can, in some cases, dilute the signature of high $N_e$ and explain
why not all accreting systems show high densities.

Having carefully distinguished between the accretion and corona in
Section~3, we reconsider the assumption that these
two components arise from different processes. G\"{u}del \& Telleschi
(2007) find that the soft X-ray excess depends both on the level of
magnetic activity as measured by the total (including coronal) $L_X$
and on the presence of accretion. While we suggest that the
role of accretion is to heat and ionize the surrounding stellar
atmosphere to create a soft X-ray (postshock cooling) plasma, it
seems possible that this plasma could be further heated to
$\sim10$~MK, either by the usual coronal heating (MHD) processes, or
possibly by the accretion energy.  Cranmer (2009) finds that in some
systems, accretion can provide the energy for coronal heating to
such high temperatures.

Studies of star forming regions have found that young stars show
strong X-ray activity, with energetic flares and extremely high X-ray
emitting temperatures; however, the increase in activity with rotation
rate breaks down for the accreting systems, leading to the conjecture
that the stellar dynamo might operate differently in these systems
(Preibisch et al.\  2005). We suggest that the accretion process may
play a more prominent role than has generally been believed. While
X-ray signatures of the accretion shock itself may not usually
dominate the emission, the energy from accretion may in fact
contribute to feeding and heating a new type of corona.

\section{Summary and Conclusions}

We have analyzed the deep \Chandra\ HETG spectrum of TW~Hya to assess
the relative contributions of accretion and coronal emission. We
summarize the most important of our findings:

\begin{itemize}

\item[1.] The spectrum shows line and continuum emission from
$\sim10$~MK coronal plasma, which may have a broad emission measure
distribution. Unlike active main sequence stars, the coronal EM
distribution abruptly cuts off below $\sim5$~MK, probably due to
absorption by the accretion stream.

\item[2.] The spectrum also shows emission lines from plasma at $T_e=
2.5$~MK, as predicted by standard models of the accretion shock. $T_e$
and $N_e$ diagnostics from He-like \ion{Ne}{9} are in excellent
agreement with accretion models for TW~Hya.

\item[3.] We measure the flux of the \ion{O}{7} forbidden line in
TW~Hya for the first time to better than 3$\sigma$, allowing a
determination of electron density.  This density is lower than $N_e$
from \ion{Ne}{9} by a factor of four and lower than $N_e$ from the
standard postshock cooling model by a factor of seven, and leads to a
new model of an accretion-fed corona.

\item[4.] The high $T_e$ ions suggest a stellar corona. By contrast,
line ratio diagnostics from the lower $T_e$ ions, indicate two regions
with different densities. In our model, the accretion shock itself has
$T_e = 3.0$~MK and $N_e=6.0 \times 10^{12}$ cm$^{-3}$, while the
postshock cooling region has $T_e = 1.75$~MK and $N_e=2.0 \times
10^{11}$ cm$^{-3}$. The postshock plasma has 30 times more mass than
the shock itself. The surface area filling factor of the shock is
1.5\% while the postshock filling factor is approximately 6.8\%.

\item[5.] Line ratio diagnostics require at least two different
absorbers, with higher column density ($N_H$) for higher charge
state. Absorption by the accretion stream is larger by a factor of 2.5
for the shock than for the postshock region.

\item[6.] A range of models for the EM distribution and elemental
abundances provides acceptable fits to the global spectrum. While
absorption, emission measure, and abundances are not independently
determined by such methods, the high Ne/O abundance seems to be a
robust result.

\item[7.] From line profile analysis of \ion{Ne}{9} lines we determine
  a subsonic turbulent velocity of about 165 km s$^{-1}$.

\end{itemize}

This high resolution X-ray spectrum of TW~Hya presents a rich set of
diagnostic emission lines for characterizing the physical conditions
of the high energy plasma and understanding the dominant physical
processes. The observations strongly support the model of an accretion
shock producing a few MK gas at high density, as first suggested by
Kastner et al.\  (2002) from an earlier HETG spectrum. The diagnostics
also support the role of accretion in producing the soft X-ray excess
at \ion{O}{7} previously discovered (G\"{u}del \& Telleschi 2007;
Robrade \& Schmitt 2007). This excess
emission requires that the accretion shock heat a large volume in the surrounding
stellar atmosphere. In TW~Hya the filling factor at \ion{O}{7} is
larger than the filling factor of the shock at \ion{Ne}{9}, and is
consistent with the filling factors of the ultraviolet and optical
continua. Both open and closed magnetic field lines may emerge from
this surrounding area to channel this ionized gas back into the
corona. While it is not yet clear whether the energy from the impact
of accretion at the star can heat a hot (10~MK) corona, accelerate a
highly ionized stellar wind, or drive hot jets, we suggest that
accretion-fed coronae may be ubiquitous in accreting young stars.

\acknowledgments

We acknowledge support from NASA to the Smithsonian Astrophysical
Observatory (SAO) under \Chandra\ GO7-8018X for GJML.  NSB and SJW
were supported by NASA contract NAS8-03060 to SAO for the Chandra
X-ray Center. SRC's contribution to this work was supported by the
Sprague Fund of the Smithonian Institution Research Endowment, and by
NASA grant NNG04GE77G to SAO. We thank Randall Smith for supporting
the customized APEC code runs. We thank the
\Chandra\ Mission Planning team for their efforts to accommodate the
ground-based observing campaign.

\end{document}